# A single differential equation description of membrane properties underlying the action potential and the axon electric field

Robert F. Melendy, Ph.D.[1,2,3]

1. Department of Electrical Engineering and Renewable Energy, The Oregon Institute of Technology, 97070 USA
2. Formerly of The School of Engineering and Computational Science, Liberty University, 24515 USA
3. E-mail any correspondence to: robert.melendy@oit.edu

## Abstract

In a succession of articles published over 65 years ago, Sir Alan Lloyd Hodgkin and Sir Andrew Fielding Huxley established what now forms our physical understanding of excitation in nerve, and how the axon conducts the action potential. They uniquely quantified the movement of ions in the nerve cell during the action potential, and demonstrated that the action potential is the result of a depolarizing event across the cell membrane. They confirmed that a complete depolarization event is followed by an abrupt increase in voltage that propagates longitudinally along the axon, accompanied by considerable increases in membrane conductance. In an elegant theoretical framework, they rigorously described fundamental properties of the Na$^+$ and K$^+$ conductances intrinsic to the action potential.

Notwithstanding the elegance of Hodgkin and Huxley's incisive and explicative series of discoveries, their model is mathematically complex, relies on no small number of stochastic factors, and has no analytical solution. Solving for the membrane action potential and the ionic currents requires integrations approximated using numerical methods. In this article I present an analytical formalism of the nerve action potential, $V_m$ and that of the accompanying cell membrane electric field, $E_m$. To conclude, I present a novel description of $V_m$ in terms of a single, nonlinear differential equation. This is an original stand-alone article: the major contribution is the latter, and how this description coincides with the cell membrane electric field. This work has necessitated unifying information from two preceding papers [1,2], each being concerned with the development of closed-form descriptions of the nerve action potential, $V_m$.

**Keywords:** Action potential; axon; core conductor; current modulation; field-dependent current; Hodgkin; Huxley; Langevin; intracellular magnetization; membrane conductance; membrane depolarization; membrane electric field, myelinated nerve; neuronal cable theory.

## I. Overview and Scope

There is abundant and well-grounded research quantifying the electrical behavior of myelinated and unmyelinated nerve fibers [3-6]. This has come to include a precise understanding of membrane impedance properties, and the longitudinal voltage and ionic currents that propagate in axon membranes [7-10] at the onset of a complete depolarization event. Of note is Hodgkin's and Huxley's quantification of ionic membrane currents and their relation to conductance and excitation in nerve [11].

Since this time, more than a few researchers have focused on rigorously describing membrane structure and electrical phenomena. A fundamental advancement in this direction was the development and prolific use of cable theory to describe signal transmission in the membrane of an axon [12-16].

In classic cable theory, axons are treated as core conducting cylinders of finite length, where the capacitive and conductance properties of the axon membrane are modeled as a distributed-parameter electrical network [17,18]. Consequently, quantitative determination of the membrane action potential and ionic currents requires solving a boundary-value problem. This approach provides a systematic means for realistically describing the action potential and the axon membrane field properties [19,20]. However, this method of modeling typically depends on the use of advanced analytical and numerical methods to solve the partial differential equations.

Comparably, the Hodgkin-Huxley equations of ionic hypothesis are a relatively complex system of differential equations that have no analytical solution: solving for the membrane action potential, membrane conductances, and the ionic currents requires integrations approximated using numerical methods.

The scope of this article is to derive an original, quantitative description of the membrane potential, $V_m$ in terms of a single, nonlinear, homogeneous differential equation. The procedure will be: (1) to present evidence that three principal factors form a basis on which the displacement of the membrane potential is described (i.e., from its resting value of $\approx$ –70 mV); (2) to synthesize these factors into a single, mathematical form for analytically computing $V_m$; (3) to demonstrate the range of phenomena to which the mathematical form is relevant. This will be achieved as follows: (*a*) Substitution of established membrane parameters into the mathematical form, followed by (*b*) computation of the membrane electric field, $E_m$; (*c*) computation of $V_m$ from its resting value through the hyperpolarizing afterpotential. These computational results will be compared with the classical findings of Hodgkin, Huxley, et al. This article will conclude with the formation of the above mentioned differential equation model.

## II. Synthesis of the Membrane Potential Analytical Model

In this section, both electrodynamic and thermodynamic evidence will be presented in forming a basis on which the displacement of the membrane potential is described. In due course, the former will be presented in a unified, analytical description of membrane excitability, followed by a description of $V_m$ in terms of a single, nonlinear, homogeneous differential equation. The basis of this description will be established in neuronal cable theory. Only certain features resulting from cable theory are of relevance to the development of the analytical model, and the applicable research will be sufficiently referenced.

### A. The Leaky Cable Conductance Property of an Axon

One solution to the neuronal cable equations is a function describing the input resistance $R_{in}$ ($\Omega$) of a leaky cable along the longitudinal length of neuronal fiber [21,22]:

$$R_{in} = R_\infty \coth X \tag{1a}$$

$$\text{s.t. } R_\infty = \frac{2d^{-3/2}}{\pi}(R_m \rho_i)^{1/2}$$

where $R_\infty$ is the input resistance of a semi-infinite cable and is proportional to the characteristic length $\lambda$ (m) of the membrane cylinder. $R_m$ represents the resistance across a unit area of membrane ($\Omega \cdot cm^2$), $\rho_i$ is the resistivity of the intracellular medium ($\Omega \cdot cm$), and $d$ is the diameter of the membrane cylinder ($\sim\mu m$). The property of intracellular resistivity is related to the axoplasmatic resistance to movement of electric charge $q$ (C) [23,24]. Extracellular resistance is considered negligible. X (Chi) is a normalized length (dimensionless). Normalized length is often given the notation "*L*" in the literature, but this is too easily confused with an actual (physical) length (m). A non-ideal X will not be constant but will vary along the length $x$ of the axon [18,23]. It is defined by $X = \int 1/\lambda \, dx$ for cylindrical membranes. This is integrated over the distances along successive (compartmental) cylindrical axes.

It is elementary to rewrite the coth term of (1a) as $R_{in} = R_\infty(1/\tanh X)$, or as $R_\infty(\cosh X /\sinh X)$. This is identical to writing $(R_{in}/\cosh X) = (R_\infty/\sinh X) = (R_\infty \operatorname{csch} X)$. Since resistance is the reciprocal of conductance $G$ ($\Omega^{-1}$), the latter may be expressed as $(1/G_{in} \cosh X)$. From this simple arrangement of terms, one can write:

$$R_\infty \operatorname{csch} X = \frac{1}{G_{in} \cosh X} \tag{1b}$$

The relevance of (1b) is it describes how a rapid drop in the input resistance of a semi-infinite cable ($R_\infty$) balances with a significant increase in the leaky cable input conductance ($G_{in}$) along the longitudinal length of neuronal fiber. Further, it's traditional and convenient to express the movement of ionic charge and polarization changes in the membrane of an axon in terms of conductance.

By and of itself, the hyperbolic conductance term (1b) is intrinsic to the displacement of the membrane potential, $V_m$ from its resting value, such that:

$$\frac{1}{G_{in} \cosh n\pi X} \propto V_m \tag{1c}$$

The inverse variation (1c) is consistent with the fact that voltage varies inversely with conductance [25]. For initial computational generality, $n\pi$ multiples of X are included in the cosh argument. The left-hand units of (1c) is Ω.

*B. Axon Intracellular Magnetization Hypothesis*

A natural consequence of a depolarizing membrane is the generation of a changing magnetic field. This is supported by a body of established research corroborating the existence of time-varying magnetic fields in an axon during the nerve impulse [26-30].

A common thread that runs through these studies is that the bioelectric activity present during the action potential produces a current in a volume conductor. For instance, the current density $J$ (A·m$^{-2}$) throughout a volume conductor generates a biomagnetic field, $B$ (T). Without exception, the latter exists in axon membranes and have been shown to be of remarkably small magnitude [27,28]. These studies offer a classical description of biomagnetic field phenomena. In contrast, what can be understood about the biomagnetic field of an axon membrane from a statistical mechanics description? Could such a description be unified with the macroscopic conductance term (1c)?

Biological tissue has been shown to have paramagnetic properties, particularly in the presence of Ca$^+$ and Na$^+$ ions [31-33]. It's therefore relevant to consider intracellular magnetization as an intrinsic membrane property and particularly, over the action potential cycle. Langevin's paramagnetic equation is suitable in this circumstance: $(M/\mu N) = \tanh(\mu B/kT)$ [34]. $M$ is magnetization (A·m$^{-1}$ or J·T$^{-1}$·m$^{-3}$), $N$ is the number of particles that make up the membrane material [with each particle having magnetic moment $\mu$ (J·T$^{-1}$)], $k$ is Boltzmann's constant (1.38 × 10$^{-23}$ J·K$^{-1}$), and $T$ is temperature (K).

Langevin's equation predicts that a paramagnetic material saturates asymptotically to the line $(M/\mu N)$ as $(\mu B/kT) \rightarrow 2$ [35]. In this instance, the fast-microscopic variables are the statistical averages of the noise generated by the thermal fluctuation of electrons in the conducting axon. The thermodynamic derivation of this noise predicts the electrical response of the axon to the resting and response potentials when the latter is quantified by the conductance. During polarization for instance, there's a considerable increase in the sodium conductance $g_{Na}$ of the axon membrane. This produces a marked increase in the current density throughout the conducting medium [36] and subsequently, an appreciable increase in the magnetization of the intracellular membrane. By Langevin's relation, it stands to reason that this intracellular magnetization saturates as all the moments become aligned against the biomagnetic field during a complete polarization event.

Based on this hypothesis, the hyperbolic conductance term (1c) and Langevin's thermodynamic relation are asserted to vary together, such that:

$$\frac{1}{\tanh\left(n\pi \frac{\mu B}{kT}\right)(G_{in} \cosh n\pi X)} \propto V_m \qquad (2)$$

The inverse variation (2) has left-hand units of Ω (since Langevin's relation is dimensionless). As with (1c), $n\pi$ multiples of $(\mu B/kT)$ are initially included in the hyperbolic argument for generality.

*C. Membrane Current Modulation Hypothesis*

An accepted and reliable method for depolarizing the excitable cells of a membrane involves variations in voltage-clamping techniques [37,38]. Regardless of method, the sensors utilized in voltage-clamping exploit the properties of the membrane potential and ionic current signals [39]. These signals are not fundamental. They're constructed of sinusoidal harmonics of the form $A \cos \omega t$, $B \sin \omega t$, or some convolution of these functions. Some signals have been shown to be unstable depending on the initial conditions in the membrane [40,41]. Irrespective of harmonics or stability, the usual practice is to quantify these signals as functions of time. The same holds true for the description of biomagnetic signals in the axon. Can the membrane current that accompanies the action potential be understood in terms of the biomagnetic *field*, i.e., $I = I(B)$?

*1. Field-Dependent Current Premise*

One can deduce *a priori* that a current $I(B)$ inevitably propagates through an axon of physical length $l$ for the period of the action potential cycle. This is perfectly reasonable, since time-varying magnetic fields have been measured in axon membranes during depolarization and hyperpolarization (as previously discussed and referenced). By Ampere's law, a field-dependent current $I(B)$ must therefore exist, such that $\oint B \cdot dl = \mu_0 I(B)$.

This prompts a fundamental question: can one quantify *variations* in $I(B)$ for the period of a depolarizing event? This would suggest the presence of a *current modulation signal*, $d^2 I(B)/dB^2$ (A·T$^{-2}$).

As is characteristic of the classically understood Na$^+$ and K$^+$ time-dependent currents, it's reasonable to assert $I(B)$ would also exhibit non-fundamental oscillatory behavior. It's well understood that a membrane response to a depolarization current pulse is accompanied by a rapid drop in the leaky cable resistance (and hence, a net increase in intracellular conductance). This necessitates a marked rate of increase in $I(B)$ during depolarization (as confirmed by Roth and Wikswo, [27]).

To mathematically synthesize a function for $I(B)$ (and to demonstrate the range to which its mathematical form is relevant), the cylindrical geometry of a classic axon [23-25] and its intrinsic electromagnetic behavior are considered basic. This is a perfectly reasonable deduction and lends itself to physical problems involving cylindrical coordinates. For instance, the description of electromagnetic fields in cavities (e.g., field strength behavior far-from and close-to cavity walls) [42,43] makes use of spherical Bessel functions, $j_n(x)$. In series notation, the spherical Bessel function is written $j_n(x) = (-1)^n x^n (x^{-1} d/dx)^n (\sin x)/x$, where $n$ is an integer (0, 1, 2, 3,..., n).

**Axiom**: The existence of a field-dependent current $I(B)$ induced in the membrane of an axon must be a response to some input excitation. Even if this excitation were an ideal impulse $\delta(x)$, the membrane could never produce a $\delta$ response (this would be physically impossible, and no experimental results have ever shown this to be the case). This would necessitate that the series $\sin(x/l)/\pi x \to \delta(x)$ in the limit as the axon length $l \to 0$ (impracticable). Since $l$ can never $\to 0$ in the limit, it follows that $I(B)$ must consists of a finite number of terms. If $I(B)$ is therefore to be modeled by a collection of spherical Bessel functions, then by the arguments made here, $I(B)$ would consist of only the first few integer values of $n$ [44].

Neurons of membranes have been shown to have natural frequency-selective feedback properties [45,46]. It stands to reason that such properties would influence how the field-strength current $I(B)$ gets transmitted, absorbed, reflected, etc. during the action potential cycle. This seems particularly true if one considers the observation of close to subcritical Hopf bifurcations in neurons, with membrane conductances and currents functioning as bifurcation parameters [41,47,48]. These phenomena support the presence of the current modulation signal, $d^2 I(B)/dB^2$.

### 2. Field-Dependent Signal Convolution Postulate

On the premise of the preceding discussion, it's reasonable to expect that $d^2 I(B)/dB^2$ would exhibit fluctuations through the membrane over the action potential cycle. This is supported by the elementary fact that a magnetic field cannot instantaneously collapse in an axon as the action potential transitions from depolarization to the hyperpolarizing afterpotential.

One plausible conjecture is that the current modulation signal behaves according to $d^2 I(B)/dB^2 \propto f(B) \otimes j_n(B)$, where $f(B)$ is some induced electromagnetic response signal and $\otimes$ is convolution. For now, it must be postulated that $f(B)$ is not a constant *and* varies nonlinearly in response to $B(t)$. Furthermore, the magnetic field must be of relatively adequate strength such that the membrane energy density (J·m$^{-3}$) is sufficient to completely depolarize the membrane.

### 3. Synthesis of the Current Modulation Function

There are chaotic nonlinearities associated with initiation of the nerve impulse by membrane depolarization [49,50]. When this is taken in conjunction with the oscillatory nature of the spherical Bessel functions $j_n(x)$ (particularly for $n$ = 0 to 2), one can reasonably hypothesize that $d^2 I(B)/dB^2$ will exhibit unstable oscillations for the period of the action potential cycle [51-53].

Without exception, unstable eigenvalues are almost always present in dynamic systems: in biological systems, there are intrinsic control mechanisms that operate in the presence of unstable equilibrium points to produce a stable response, often *after* a margin of instability [40,54-56]. On the premise of unstable oscillations, the simplest case would be a signal quantified by $(-1)^n x^n (x^{-1} d/dx)^n \sin x/x \otimes f(B)$ for $n = 0$ and for nonlinear $f(B) = x^2$. On that account, it follows that $j_0(x) \otimes x^2 \equiv x \sin x$. Hence, the initial prediction is that $d^2 I(B)/dB^2 = (2\pi a/\mu_0) \times (t \sin n\pi t)$, where the inclusion of $(2\pi a/\mu_0)$ is a consequence of Ampere's law. The axon radius is $a$ ($\sim\mu$m) and $\mu_0$ is the vacuum permeability of free space ($4\pi \times 10^{-7}$ H·m$^{-1}$).

### 4. An Initial Quantitative Description of the Action Potential

The question posed – *Can the membrane current that accompanies the action potential be quantified in terms of the biomagnetic field $I(B)$?* – can now be addressed.

By Ohm's law, $V \propto I$. Hence, the displacement of the membrane potential $V_m$ must be in proportional-variation to $I(B)$, as well as any $n^{th}$ derivative of $I(B)$, such that:

$$V_m \propto \left(\frac{2\pi a}{\mu_0}\right)(t \sin n\pi t) \times \frac{1}{\tanh\left(n\pi \frac{\mu B}{kT}\right)(G_{in}\cosh n\pi X)} \quad (3a)$$

The right-hand units of (3a) are V·T$^{-2}$. As before, $n$-multiples of $\pi$ are initially built-into the transcendentals. For (3a) to have units of V, the right-hand side must be multiplied by the square of the membrane magnetic field, $B_m$ (T):

$$V_m = \frac{B_m^2 \left(\frac{2\pi a}{\mu_0}\right)(t \sin n\pi t)}{\tanh\left(n\pi \frac{\mu B}{kT}\right)(G_{in}\cosh n\pi X)} \quad (3b)$$

(3b) has units of V and offers an initial analytical description of the membrane action potential $V_m$. As per the scope of this article, the next step will be to express (3b) in terms of the accompanying cell membrane electric field, $E_m$.

### III. The Membrane Electric Field Hypothesis

The numerator of (3b) is a current term having units of amps (A). Electrical current in the axon per unit area of axon cross section is $J = I/A$ (A·m$^{-2}$), where $A = \pi a^2$. This current density may also be described as $J = E_m/\rho_m$, where $E_m$ is the axon membrane electric field (V·m$^{-1}$) and $\rho_m$ is the longitudinal membrane resistivity (Ω·m). Hence, the axon current flow may be written as $I = JA = (\pi a^2)E_m/\rho_m$. By the laws of classical electrodynamics [57], the numerator of (3b) may therefore be alternatively expressed as

$$B_m^2\left(\frac{2\pi a}{\mu_0}\right)(t\sin n\pi t) = E_m\left(\pi a^2\right)\rho_m^{-1}(t\sin n\pi t) \quad (3c)$$

The electric field is considered constant along the axon longitudinal axis but is radially-dependent, such that **E$_m$** = $E_m\hat{\mathbf{u}}_r$, where $\hat{\mathbf{u}}_r$ is a unit vector in the axon radial direction. It's more practical therefore to express the right-hand side of (3c) in terms of the axon thickness $\Delta r$ [18, 23-25], such that $E_m\hat{\mathbf{u}}_r \propto \Delta r E_m$.

Consider now the introduction of a proportionality constant $k$, such that $k\Delta r E_m$ produces units of amps (A). Then $k$ would need to have units of (F·m$^{-1}$) × (V·m$^{-1}$). The conjecture therefore is that $k = \varepsilon_0 E_m$, where $\varepsilon_0$ is the vacuum permittivity of free space (8.854 × 10$^{-12}$ F·m$^{-1}$). Substituting these relations into (3b) gives:

$$V_m = \frac{\varepsilon_0 \Delta r E_m^2 (t\sin n\pi t)}{\tanh\left(n\pi\dfrac{\mu B}{kT}\right)(G_{in}\cosh n\pi X)} \quad (3d)$$

The units of volts are preserved in going from (3b) to (3d). It will be subsequently shown that (3d) gives a correct description of the classic nerve action potential and the cell membrane electric field. Computational results for $V_m$ and $E_m$ will be validated by comparison with standardized values in the literature.

### IV. Materials and Methods

A `Matlab` algorithm was developed to computationally test the modeling suitability of (3d). This required a practical choice of physical membrane parameters [14,23,25,35,36]:

| | |
|---|---|
| Axon thickness (myelinated): | $\Delta r$ = 2 μm |
| Axon ("cable") length: | $0 \leq x \leq 4000$ μm |
| Length constant: | $\lambda$ = 1000 μm |
| Resistance (unit area of membrane): | $R_m$ = 2.56 Ω·m$^2$ |
| Intracellular resistivity: | $\rho_i$ = 0.4 Ω·m |
| Input resistance (semi ∞ cable): | $R_\infty$ = 20.3718(32) MΩ |
| Nonlinear magnetization (unitless): | $0 \leq (\mu B/kT) \leq 4$ |
| Action potential cycle time: | $0 \leq t \leq 5$ msec |
| Vacuum permittivity $\varepsilon_0$: | 8.854(10$^{-12}$) F/m |

#### A. Parameterizing the Conductance Term

The `Matlab` algorithm was used to first compute an appropriate number of $n\pi$ multiples for each of the transcendental arguments of (3d). For the cosh argument, a best-fit iteration returned a value of $n \approx 1$.

#### B. Parameterizing the Magnetization Term

The magnetization factor tanh ($n\pi\mu B/kT$) was not providing the parameterization to correctly model the unique dynamics of a classical membrane action potential cycle.

An asymptotic series expansion [58] of this factor was performed to reveal the sensitivities associated with each of the terms in the tanh argument. In consequence, it was established that this factor was best-fit to an exponential function having the form $b^{\tanh(n\pi\mu B/kT)}$.

A first prediction for $b$ was the natural exponential $e$, but this produced a neuronal firing-delay and bursting effect not characteristic of classical membrane action potential cycles from the literature. The `Matlab` algorithm was coded to converge on a suitable estimate: $1.414 \leq b \leq 1.732$, producing an average estimate of $b \approx \pi/2$. For the $n\pi$ multiple, the algorithm returned a best-fit iteration of $n \approx 4$.

#### C. Parameterizing the Field Current Term

For the sin argument of (3d), the Matlab algorithm returned a best-fit iteration of $n \approx 1$. However, the simulation kept producing an abnormally-shaped action potential cycle. It was suspected that the sin term was displaying a sensitivity-dependence on the initial conditions.

This reasoning supported the notion of *Lyapunov's stability criterion*, and the possible need for computing a Lyapunov characteristic number, $\xi$ [59]. The Lyapunov characteristic number provides information about the rate-of-separation of infinitesimally close trajectories. Classically, $\lambda$ is used for the Lyapunov characteristic number, but $\lambda$ has been used in this article for the axon length constant (hence, $\xi$ was chosen).

A double-precision floating point `Matlab` algorithm was written to compute $\xi$ using the chosen physical parameters, resulting in the estimate $\xi = 1.357(50) \approx e/2$. The pure fact that $\xi > 0$ was not surprising since the field current signal was predicted to exhibit unstable oscillations during membrane depolarization (SEE SECTION II, C.3). (3d) was amended to account for the sensitivity-dependence of the field current signal:

$$V_m = \frac{\varepsilon_0 \Delta r E_m^2 \left(t^{0.5e}\sin n\pi t\right)}{(0.5\pi)^{\tanh\left(4\pi\frac{\mu B}{kT}\right)}(G_{in}\cosh \pi X)} + d \quad (4a)$$

Where the intracellular resting potential of the membrane $d$ (relative to the outside of the cell) is also accounted for.

The `Matlab` algorithm was used to establish an estimate of $\varepsilon_0 \Delta r E_m^2$ to a value necessary to initiate complete depolarization of the axon, converging to a value of 2.070(39) × 10$^{-8}$. If all previously discussed hypotheses are sound, this estimate will have the units V$^2$·F·m$^{-2}$.

### D. Ethical Approval
The conducted research is not related to either human or animal use.

### V. Results: Reliability of the Hypothesized Model (4a)
To demonstrate that (4a) gives a correct description of the classic nerve action potential and the cell membrane electric field, computational results for $V_m$ and $E_m$ are validated next.

### A. Confirming the Membrane Electric Field, $E_m$
The thickness of a myelinated cell membrane is $\Delta r \approx 2$ μm [23,60,61]. It follows that: $\varepsilon_0 \Delta r E_m^2 =$ (8.854 × 10$^{-12}$ F·m$^{-1}$) × (2 μm) × $E_m^2 =$ 2.070(39) × 10$^{-8}$ V$^2$·F·m$^{-2}$. This results in $E_m =$ 3.418(95) × 10$^4$ V·m$^{-1}$.

A classic axon membrane model will have a potential difference between the interior and exterior side of the membrane of $\Delta V_m \approx -70$ mV [62]. The theoretical electric field for a myelinated membrane of 2 μm thickness is therefore $E_m = -dV_m/d(\Delta r) = -(-70$ mV$)/(2$ μm$) = 3.5 \times 10^4$ V·m$^{-1}$ [23,60,63]. This theoretical result is highly consistent with the computation of the electric field from the analytical model (4a), having a percent error ≈ 2.3%. This is an initial confirmation that (4a) is a correct description of the classical membrane action potential cycle, $V_m$.

### B. Confirming the Membrane Action Potential, $V_m$
To further establish that (4a) provides a correct description of a classical action potential cycle, a computational profile of $V_m$ was completed for $0 \leq t \leq 5$ msec.

Compiling all preceding factors into the `Matlab` algorithm gives a restoration voltage of $d \approx -67.9$ mV. Hence for $t < 0$, $V_m = -67.9$ mV. For $0 \leq t \leq 5$ msec:

$$V_m = \frac{\varepsilon_0 \Delta r E_m^2 \left(t^{0.5e} \sin \pi t\right)}{(0.5\pi)^{\tanh\left(4\pi \frac{\mu B}{kT}\right)} (G_{in} \cosh \pi X)} - 67.9 \times 10^{-3} \text{ V} \quad \textbf{(4b)}$$

Figure 1 is a plot of $V_m$ vs. $t$ from **(4b)**† and demonstrates the classical action potential voltage cycle in nerve under stable equilibrium conditions [11,18,23,25,61,63].

### VI. Discussion

### A. Inference of Ionic Current Flow
In the classical Hodgkin-Huxley model, it's well-known that the lipid bilayer of the axon membrane is modeled as a lumped-capacitance $C_m$ (F) [11,64], such that $C_m = \varepsilon_0 \Delta r$.

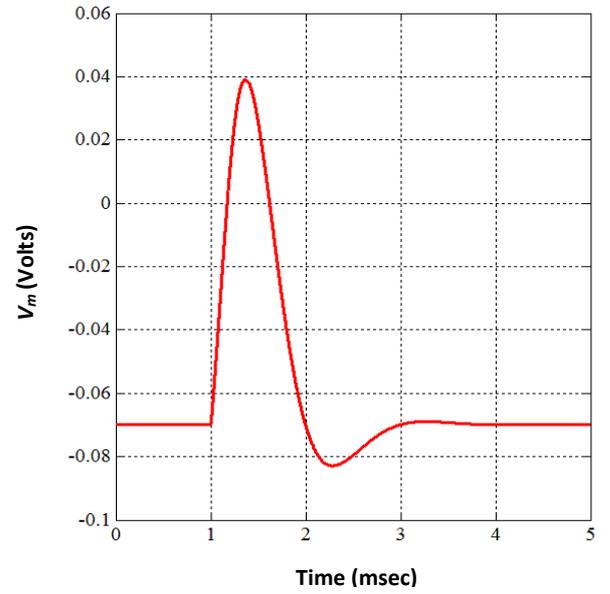

**Fig.1:** The completely depolarized membrane potential **(4b)** exhibits stable equilibrium for $\xi = 0.5e$. †This plot was time-shifted to the right by 1 ms (intentioanlly) to visually enhance the action potential cycle.

Also well-known to the Hodgkin-Huxley model is the quantification of the ionic current flow $I_C$ through this bilayer. In terms of $C_m$, the current is described as $I_C = C_m \times (dV_m/dt)$, where $V_m$ is the membrane action potential.

The relationship between the membrane electric and magnetic fields $E_m$ and $B_m$ may be expressed in terms of the time rate-of-change of the membrane potential, such that $dV_m/dt = E_m^2/B_m$. The current flow through the membrane lipid bilayer may likewise be expressed in terms of these fields: $C_m \times (dV_m/dt) = C_m \times (E_m^2/B_m)$. Thus, a novel feature of **(4b)** is that of offering an alternative description to the classical model for the time-dependence of the membrane current $I_C$ in terms of $E_m$.

The field current term in **(4b)** has units of amps, where $\varepsilon_0 \Delta r E_m^2$ has units of V$^2$·F·m$^{-2}$: but this is also units of T$^2$. The point is that $\varepsilon_0 \Delta r E_m^2 \left(t^{0.5e} \sin \pi t\right) = B_m^2 \left[d^2 I(B)/dB^2\right]$, meaning once again that the time-dependent current $I_C$ of the Hodgkin-Huxley model is inferred here by the electric field-induced current term of **(4b)**.

It is hoped this makes clear the implicit manifestation of the membrane current underlying the action potential and its relationship to $E_m$. The concluding hypothesis is that **(4b)** resolves the biophysics of how an axon conducts the action potential cycle in a single analytical description.

### B. Formation of the Differential Equation Description
From **(4b)**, let:

$$\Gamma = \frac{\varepsilon_0 \Delta r \left(t^{0.5e} \sin \pi t\right)}{(0.5\pi)^{\tanh\left(4\pi \frac{\mu B}{kT}\right)} (G_{in} \cosh \pi X)} \quad \textbf{(4c)}$$

So then $V_m = \Gamma E_m^2 - 67.9 \times 10^{-3} \Rightarrow E_m^2 = \Gamma^{-1}(V_m + 67.9 \times 10^{-3})$.

From the laws of electrodynamics [57], $E_m = -\nabla V_m$ ($\nabla$ = del operator). Hence, $E_m^2 = (-\nabla V_m)^2 = (\nabla V_m)^2$ or:

$$(\nabla V_m)^2 - \frac{1}{\Gamma}(V_m + 67.9 \times 10^{-3}) = 0 \quad \text{(4d)}$$

As previously mentioned, the electric field is constant along the axon's longitudinal axis, but is radial-dependent such that $\mathbf{E}_m = E_m \hat{\mathbf{u}}_r$ (refer to SECTION III). Hence, $\nabla V_m$ in (4d) reduces to $= \partial V_m / \partial(\Delta r)$. Consequently:

$$\left[\frac{\partial V_m}{\partial(\Delta r)}\right]^2 - \frac{1}{\Gamma}V_m - u = 0 \quad \text{(4e)}$$

Where $u = (67.9 \times 10^{-3})\Gamma^{-1}$. Equation **(4e)** is significant in that it offers a novel rigorous quantification of the membrane action potential in closed-form. The traditional Hodgkin-Huxley quantification of the membrane potential requires numerically integrating four differential equations to solve for $V_m$ and to this discussion, I have nothing to add.

### VII. Summary

The development of an original, quantitative description of the membrane (action) potential displacement $V_m$ was presented in this article. This description is a conductance-based model rooted in cable theory. Unlike the traditional Hodgkin-Huxley equations of ionic hypothesis, I did not explicitly describe the action potential in the context of ion channels (i.e., the chemistry and physics behind the contribution of different ions to the action potential cycle are not explicit or necessary features of my model).

1. Evidence was given that three principal factors form a basis on which the membrane potential displacement is described. These three factors are the axon leaky cable conductance, intracellular membrane magnetization, and membrane current modulation.

2. These three hypothesized factors were unified in a single analytical form for quantitatively determining $V_m$.

3. Beginning with substitution of established membrane parameters, the range of phenomena to which the analytical form is relevant was demonstrated by: (*a*) computation of the membrane electric field, $E_m$; (*b*) computation of the membrane potential cycle, $V_m$.

4. One of the novelties of this work is that it provides a mechanistic understanding of how intracellular conductance, the thermodynamics of magnetization, and current modulation function together to generate excitation in nerve.

5. Another novel feature of this work is the statistical mechanics description of intracellular magnetization, and how this phenomenon relates to the presence of ions in the membrane channel.

6. The significance of this model is that it offers an original and fundamental advancement in the understanding of the action potential in a unified analytical description. It provides a conductive, thermodynamic, and electro-magnetic explanation of how an action potential propagates in nerve in a single mathematical construct.

7. Another significant feature of this model is that it offers a new and rigorous description of the action potential, quantified as a single, nonlinear differential equation in $V_m$. This is in contrast to the traditional Hodgkin-Huxley equations of ionic hypothesis, which consists of four differential equations having no analytical solution.

### Compliance with Ethical Standards

Conflict of Interests: The author Robert F. Melendy, Ph.D. declares that I have no conflict of interest(s).

### References

1. R.F. Melendy, Resolving the biophysics of axon transmembrane polarization in a single closed-form description. Journal of Applied Physics, 118(24), (2015). https://doi.org/10.1063/1.4939278

2. R.F. Melendy, A subsequent closed-form description of propagated signaling phenomena in the membrane of an axon. AIP Advances, 6(5), (2016). https://doi.org/10.1063/1.4948985

3. A.L. Hodgkin, Evidence for electrical transmission in nerve. Journal of Physiology, 90, 183-210 (1937). https://doi.org/10.1113/jphysiol.1937.sp003507

4. J.B. Hursh, Conduction velocity and diameter of nerve fibers. American Journal of Physiology, 127, 131-139 (1939). https://doi.org/10.1152/ajplegacy.1939.127.1.131

5. B. Frankenhaeuser, The ionic currents in the myelinated nerve fiber. Journal of General Physiology, 48, 79-81 (1965). https://doi.org/10.1085/jgp.48.5.79

6. B. Naundorf, F. Wolf, M. Volgushev, Unique features of action potential initiation in cortical neurons. Nature, 440, 1060-1063 (2006). https://doi.org/10.1038/nature04610

7. K.S. Cole, H.J. Curtis, Electric impedance of the squid giant axon during activity. Journal of General Physiology, 22, 649-670 (1939). https://doi.org/10.1085/jgp.22.5.649

8. D.E. Goldman, Potential, impedance, and rectification in membranes. Journal of General Physiology, 27, 37-60 (1943). https://doi.org/10.1085/jgp.27.1.37

9. A.L. Hodgkin, B. Katz, The effect of sodium ions on the electrical activity of the giant axon of the squid. Journal of Physiology, 108, 37-77 (1949). https://doi.org/10.1113/jphysiol.1949.sp004310

10. J. Koester, S.A. Siegelbaum, in Principles of Neural Science, E.R. Kandel, J.H. Schwartz, T.M. Jessell, Eds. (McGraw-Hill, New York, 2000), pp. 140-149.

11. A.L. Hodgkin, A.F. Huxley, A quantitative description of membrane current and its application to conduction and excitation in nerve. Journal of Physiology, 117, 500-544 (1952). https://doi.org/10.1113/jphysiol.1952.sp004764

12. R.E. Taylor, in Physical Techniques in Biological Research, W.L. Natsiik, Ed. (Academic Press, New York, 1963), pp. 219-262.

13. R. Iansek, S.J. Redman, An analysis of the cable properties of spinal motoneurones using a brief intracellular current pulse. Journal of Physiology, 234, 613-636 (1973). https://doi.org/10.1113/jphysiol.1973.sp010364

14. W. Rall, J. Segev, The Theoretical Foundation of Dendritic Function: Selected Papers of Wilfrid Rall with Commentaries (MIT Press, Boston, MA, 1995).

15. M. London, C. Meunier, I. Segev, Signal transfer in passive dendrites with nonuniform membrane conductance. Journal of Neuroscience, 19, 8219-8233 (1999). https://doi.org/10.1523/JNEUROSCI.19-19-08219.1999

16. F. Nadim, J. Golowasch, Signal transmission between gap-junctionally coupled passive cables is most effective at an optimal diameter. Journal of Neurophysiology, 95, 3831-3843 (2006). https://doi.org/10.1152/jn.00033.2006

17. H.M. Lieberstein, On the Hodgkin-Huxley partial differential equation. Mathematical Biosciences, 1, 45-69 (1967). https://doi.org/10.1016/0025-5564(67)90026-0

18. W. Rall, Core Conductor Theory and Cable Properties of Neurons: Handbook of Physiology, the Nervous System, Cellular Biology of Neurons (American Physiological Society, 1977), pp. 39-93.

19. R. West, E. Schutter, G. Wilcox, in The IMA Volumes in Mathematics and its Applications: Evolutionary Algorithms, L.D. Davis et al., Eds. (Springer, New York, 1999), pp. 33-64.

20. C. Bédard, A. Destexhe, A modified cable formalism for modeling neuronal membranes at high frequencies. Biophysical Journal, 94, 1133-1143 (2008). https://doi.org/10.1529/biophysj.107.113571

21. J.J.B. Jack, D. Noble, R.W. Tsien, Electric Current Flow in Excitable Cells (Clarendon Press, Oxford, 1975).

22. D. Sterratt, Principles of Computational Modelling in Neuroscience (Cambridge University Press, Cambridge, 2011). https://doi.org/10.1017/CBO9780511975899

23. R. Hobbie, Intermediate Physics for Medicine and Biology (AIP Press, New York, 1997).

24. R. Plonsey, R. Barr, Bioelectricity: A Quantitative Approach (Springer, Boston, 2000). https://doi.org/10.1007/978-1-4757-3152-1

25. N. Sperelakis, N. Sperelakis, Cell Physiology Sourcebook: Essentials of Membrane Biophysics (Academic Press, London, 2012).

26. J. Malmivuo, R. Plonsey, Bioelectromagnetism: Principles and Applications of Bioelectric and Biomagnetic Fields (Oxford University Press, New York, 2000).

27. B. Roth, J. Wikswo, The magnetic field of a single axon: a comparison of theory and experiment. Biophysical Journal, 48, 93-109 (1985). https://doi.org/10.1016/S0006-3495(85)83763-2

28. B. Roth, J. Wikswo, The electrical potential and the magnetic field of an axon in a nerve bundle. Mathematical Biosciences, 76, 37-57 (1985). https://doi.org/10.1016/0025-5564(85)90045-8

29. R.S. Wijesinghe, Detection of magnetic fields created by biological tissues. Journal of Electrical and Electronic Systems, 3, 1-7 (2014). https://doi.org/10.4172/2332-0796.1000120

30. B. Greenebaum, F. Barnes, Bioengineering and Biophysical Aspects of Electromagnetic Fields (CRC/Taylor & Francis, Boca Raton, FL., 2007).

31. B. Commoner, J. Townsend, G.E. Pake, Free radicals in biological materials. Nature, 174, 689-691 (1954). https://doi.org/10.1038/174689a0

32. V.N. Varfolomeev et al., Paramagnetic properties of hepatic tissues and transplantable hepatomas. Biofizika. 21, 881-886 (1976).

33. R. Pethig, D.B. Kell, The passive electrical properties of biological systems: their significance in physiology, biophysics and biotechnology. Physics in medicine and biology, 32, 933-970 (1987). https://doi.org/10.1088/0031-9155/32/8/001

34. C. Kittel, Introduction to Solid State Physics (Wiley, New York, 2008).

35. W.T. Coffey, Y.P. Kalmykov, J.T. Waldron, The Langevin Equation, with Applications in Physics, Chemistry, and Electrical Engineering (World Scientific, River Edge, NJ, 1996).

36. J. Koester, S.A. Siegelbaum, in Principles of Neural Science, E.R. Kandel, J.H. Schwartz, T.M. Jessell, Eds. (McGraw-Hill, New York, 2000), pp. 150-169.

37. A.F. Huxley, From overshoot to voltage clamp. Trends in Neurosciences, 25, 553-558 (2002). https://doi.org/10.1016/S0166-2236(02)02280-4

38. E.O. Hernández-Ochoa, M.F. Schneider, Voltage clamp methods for the study of membrane currents and SR Ca2+ release in adult skeletal muscle fibres. Progress in Biophysics and Molecular Biology, 108, 98-118 (2012). https://doi.org/10.1016/j.pbiomolbio.2012.01.001

39. S.G. Waxman, J.D. Kocsis, P.K. Stys, Eds., The Axon: Structure, Function and Pathophysiology (Oxford University Press, New York, 1995). https://doi.org/10.1093/acprof:oso/9780195082937.001.0001

40. A.V. Holden, P.G. Haydon, W. Winlow, Multiple equilibria and exotic behavior in excitable membranes. Biological Cybernetics, 46, 167-172 (1983). https://doi.org/10.1007/BF00336798

41. 41. R. Guttman, S. Lewis, J. Rinzel, Control of repetitive firing in squid axon membrane as a model for a nuroneoscillator. Journal of Physiology, 305, 377-395 (1980). https://doi.org/10.1113/jphysiol.1980.sp013370

42. H.R. Leuchtag, Voltage-Sensitive Ion Channels: Biophysics of Molecular Excitability (Springer, New York, Philadelphia, 2008). https://doi.org/10.1007/978-1-4020-5525-6

43. D.A. Hill, Electromagnetic Fields in Cavities: Deterministic and Statistical Theories (IEEE Press Series on Electromagnetic Wave Theory, NJ, 2009). https://doi.org/10.1002/9780470495056

44. D.A. McQuarrie, Mathematical Methods for Scientists and Engineers (University Science Books, CA, 2003).
45. R. FitzHugh, Impulses and physiological states in theoretical models of nerve membrane. Biophysical Journal, 1, 445-466 (1961). https://doi.org/10.1016/S0006-3495(61)86902-6
46. G. Zhao, Z. Hou, H. Xin, Frequency-selective response of FitzHugh-Nagumo neuron networks via changing random edges. Chaos: An Interdisciplinary Journal of Nonlinear Science, 16, 043107 (2006). https://doi.org/10.1063/1.2360503
47. S.Y. Gordleeva, et al., Bi-directional astrocytic regulation of neuronal activity within a network. Frontiers in Computational Neuroscience, 6, 104-114 (2012). https://doi.org/10.3389/fncom.2012.00092
48. R.W. Aldrich, P.A. Getting, S.H. Thompson, Inactivation of delayed outward current in molluscan neurone somata. Journal of Physiology, 291, 507-530 (1979). https://doi.org/10.1113/jphysiol.1979.sp012828
49. K. Aihara, G. Matsumoto, in Nerve Excitation and Chaos: Dynamical Systems and Nonlinear Oscillations, Gikō Ikegami, Ed. (World Scientific Publishing Co., 1986). Pp. 254-267.
50. J. Rinzel, G. Huguet, Nonlinear Dynamics of Neuronal Excitability, Oscillations, and Coincidence Direction. Communications on Pure and Applied Mathematics, 66(9), 1464-1494 (2013). https://doi.org/10.1002/cpa.21469
51. Morris, H. Lecar, Voltage oscillations in the barnacle giant muscle fiber. Biophysical Journal, 35, 193-213 (1981). https://doi.org/10.1016/S0006-3495(81)84782-0
52. T. Sasaki, N. Matsuki, Y. Ikegaya, Action-potential modulation during axonal conduction. Science, 331, 599-601 (2011). https://doi.org/10.1126/science.1197598
53. 53. N.H. Sabah, K.N. Leibovic, The effect of membrane parameters on the properties of the nerve impulse. Biophysical Journal, 12, 1132-1144 (1972). https://doi.org/10.1016/S0006-3495(72)86150-2
54. N.F. Britton, Essential Mathematical Biology (Springer-Verlag, London, 2003). https://doi.org/10.1007/978-1-4471-0049-2
55. J.D. Murray, Mathematical Biology I: An Introduction (Springer-Verlag, Berlin, 2002).
56. E.O. Voit, A First Course in Systems Biology (Garland Science, Taylor & Francis, New York, 2013).
57. R.L. Armstrong, J.D. King, The Electromagnetic Interaction (Prentice Hall, Englewood Cliffs, NJ, 1973).
58. G.B. Arfken, H.J. Weber, F.E. Harris, Mathematical Methods for Physicist: A Comprehensive Guide (Elsevier, MA, 2013).
59. E. Weisstein, CRC Concise Encyclopedia of Mathematics (CRC Press, Boca Raton, 2003).
60. The electrical system of the body: The physics of the nervous system (Medical Physics, University of Notre Dame, n.d., http://www3.nd.edu/~nsl/Lectures/mphysics/).
61. R.I. Macey, in Membrane Physiology, T.E. Andreoli, J.F. Hoffman, D.D. Fanestil, Eds. (Springer, New York, 1980), pp. 125-146. https://doi.org/10.1007/978-1-4757-1718-1_7
62. T. Begenisic, Magnitude and location of surface charges on myxicola giant axons. The Journal of General Physiology, 66, 47-65 (1975). https://doi.org/10.1085/jgp.66.1.47
63. J. Enderle, S. Blanchard, J. Bronzino, Introduction to Biomedical Engineering (Elsevier Academic Press, Amsterdam, Boston, London, New York, 2005).
64. P. Smejtek, in Permeability and Stability of Lipid Bilayers, E. Anibal Disalvo, S.A. Simon, Eds. (CRC Press, Boca Raton, Ann Arbor, London, 1994), pp. 197-236.